\documentclass[conference]{IEEEtran}
\IEEEoverridecommandlockouts
\usepackage{cite}
\usepackage{amsmath,amssymb,amsfonts}
\usepackage{algorithmic}
\usepackage{graphicx}
\usepackage{textcomp}
\usepackage{xcolor}
\usepackage{subfig}

\def\BibTeX{{\rm B\kern-.05em{\sc i\kern-.025em b}\kern-.08em
    T\kern-.1667em\lower.7ex\hbox{E}\kern-.125emX}}

\begin{document}
\bstctlcite{IEEEexample:BSTcontrol}

\title{Demo: Real-Time Semantic Communications with a Vision Transformer}

\author{
    \IEEEauthorblockN{Hanju Yoo,$^{1}$ Taehun Jung,$^{1}$ Linglong Dai,$^{2}$ Songkuk Kim,$^{1}$ and Chan-Byoung Chae$^{1}$}
    \IEEEauthorblockA{$^{1}$School of Integrated Technology, Yonsei University, Korea\\ $^{2}$Department of Electronic Engineering, Tsinghua University, China
    \\\{hanju.yoo, taehun.jung, songkuk, cbchae\}@yonsei.ac.kr, daill@tsinghua.edu}
}

\maketitle

\begin{abstract}
Semantic communications are expected to enable the more effective delivery of meaning rather than a precise transfer of symbols. In this paper, we propose an end-to-end deep neural network-based architecture for image transmission and demonstrate its feasibility in a real-time wireless channel by implementing a prototype based on a field-programmable gate array (FPGA). We demonstrate that this system outperforms the traditional 256-quadrature amplitude modulation system in the low signal-to-noise ratio regime with the popular CIFAR-10 dataset. To the best of our knowledge, this is the first work that implements and investigates real-time semantic communications with a vision transformer.

\end{abstract}

\begin{IEEEkeywords}
Semantic communications, image transmission, real-time wireless communications, deep neural network
\end{IEEEkeywords}

\section{Introduction}

With the emergence of immersive contents, such as extended reality (XR), communication systems must simultaneously achieve high data rates and low latencies. These contradictory requirements have led to ongoing races for wider bandwidths and higher costs. As a result of the rapid advances in computing power and more data, researchers have been attempting to improve the spectral efficiency by introducing deep neural networks (DNNs) to communications to enable the faster and more accurate transmission of symbols with limited bandwidths.

What matters, however, is the effective delivery of meaning to achieve goals, not the bits. Some early research regarding semantic communications has been conducted in the scope of text~\cite{reinforcementlearning} or speech transmission~\cite{deepsc-speech}. Nevertheless, none of the research has addressed the image transmission problem, which is the primary source of human information perception and accounts for most traffic usage on the internet. Furthermore, their results were validated only in simulated environments and did not equalize the system data rate for a fair comparison.

In this paper, we introduce an end-to-end DNN-based communication system for images and implement a real-time wireless image transmission platform based on a field-programmable gate array to demonstrate the feasibility of the system on actual channel conditions. We compare the proposed method to the traditional 256-quadrature amplitude modulation (QAM) system to demonstrate its superiority in the low signal-to-noise (SNR) regime.

\begin{figure*}[htbp]
	\centering{\includegraphics[width=\textwidth,height=4.2cm]{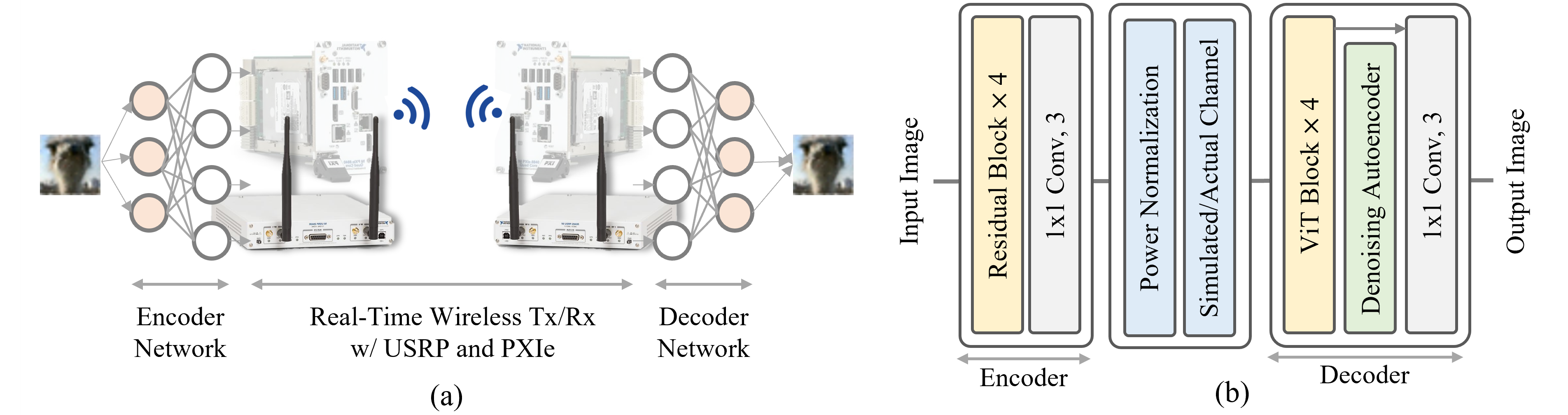}}
	\caption{(a) System setup and (b) proposed deep neural network (DNN) system architecture.}
	\label{fig:sysarch}
\end{figure*}

\begin{figure*}[htbp]
	\centering
	\includegraphics[width=\textwidth,height=3.6cm]{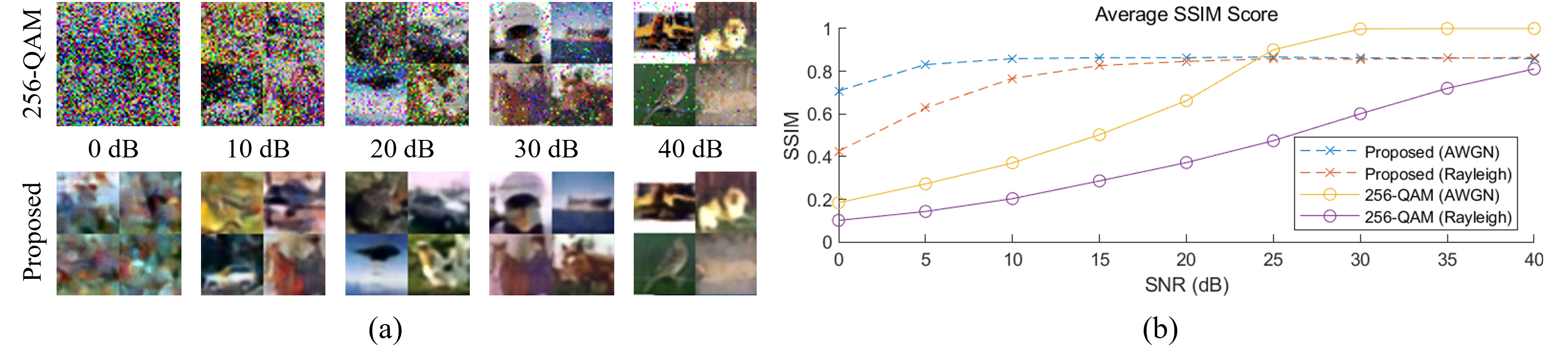}
	\caption{(a) Transmitted images and (b) structural similarity index measure (SSIM) results of the proposed and baseline systems.}
	\label{fig:results-summary}
\end{figure*}

\section{System Setup} \label{systemsetup}

\subsection{System Architecture} \label{sysarch}

As presented in Figure~\ref{fig:sysarch}, the system consists of three blocks: an image encoder, channel layer, and image reconstruction block (decoder).  The encoder and decoder are trainable neural networks whereas the channel layer is either a simulated (for training) or physical wireless channel based on a software-defined radio (for demonstration).

\par\smallskip
\noindent\textbf{Image encoder.}
The encoder network $F_{\text{encode}}$ consists of four consecutive $3\times3$ convolutional layers that follow the prominent bottleneck residual block~\cite{resnet} design, known for good early visual processing and feature extraction performance. Given an image sequence $s \in \mathbb{R}^{H \times W \times C}$ (where $H, W$, and $C$ are the image height, width, and channel, respectively), the encoder network $F_{\text{encode}}$ transforms $s$ into symbols $x \in \mathbb{R}^{H \times W \times C}$, denoted as follows:
\begin{equation} \label{encoder_eq}
x = F_{\text{encode}}(s).
\end{equation}

\par\smallskip
\noindent\textbf{Channel layer.}
The channel layer conducts two main tasks: power normalization and physical channel simulation. In this block, the symbol $x$ is divided by the power normalizer coefficient $c$ to satisfy the power constraint $\mathop{\mathbb{E}}\| x \|^2 = 1.$ Then, the normalized symbol $\bar{x} = x / c$ is transmitted via a communication link, which can be modeled as follows:
\begin{equation} \label{channel_eq}
y = h \bar{x} + n,
\end{equation}
with a channel coefficient $h$ and an additive Gaussian noise $n \sim \mathcal{N}(0,\,\sigma^{2})$ with noise variance $\sigma^{2}$.

\par\smallskip
\noindent\textbf{Image reconstruction block.}
The transmitted symbol $y$ is transformed to produce the naively decoded symbol ${\hat{x} = c \cdot y/h}$. We assumed that the perfect channel source information and the power normalization coefficient are known. Then, the decoder network $F_{\text{decode}}$ reconstructs the image sequence $\hat{s}$ from $\hat{x}$. The decoder network consists of a vision transformer (ViT)~\cite{ViTpaper} followed by denoising autoencoder (DAE)~\cite{denoisingae} layers. Vision transformer is known for useful properties such as resilience to typical image occlusions, and DAE effectively denoises images by extracting critical features from the corrupted image. A shortcut between two layers is introduced to facilitate the recovery of detailed textures, which may be lost on denoising.
The overall decoder operation can be expressed by the following:
\begin{equation} \label{vit_eq}
x_{0} = F_{\text{vit}}(\hat{x}),
\end{equation}
\begin{equation} \label{decoder_eq}
\hat{s} = W x_{0} + F_{\text{dae}}(x_{0}),
\end{equation}
where $\hat{s}$ denotes the reconstructed image sequence, and $W$ is a trainable matrix. Moreover, $F_{\text{vit}}$ and $F_{\text{dae}}$ are the ViT and DAE blocks, respectively.

\subsection {Experiments} \label{experiments}

We trained the system with a CIFAR-10 image dataset containing 50,000 training and 10,000 test images. The total parameter count of the system was 0.76M. We trained the network for 150 epochs with a Rayleigh channel at 35~dB SNR and with the pixel-wise mean squared error loss. Then we fine-tuned the system for another 150 epochs at 15~dB with the mean absolute error loss. The Adam optimizer with $\ell$$r=0.002$ was adopted for parameter optimization.

We evaluated the model on two channel models, additive white Gaussian noise and Rayleigh, with a SNR ranging from 0 to 40~dB. A 256-QAM system with average symbol power normalization was selected for the baseline, as the proposed system sends one symbol per image pixel (which can have 256 levels).\footnote{The 256-QAM system does not have sophisticated coding techniques. Future work will consider various source and/or channel coding schemes.} We used the structural similarity index (SSIM)~\cite{ssim} to evaluate the system performance.

\subsection {Results} \label{results}

Figure~\ref{fig:results-summary} illustrates the transmitted images and their SSIM scores for the proposed and baseline 256-QAM systems. For Figure~\ref{fig:results-summary}(a), we assumed a Rayleigh channel and sent random images to produce the results. For Figure~\ref{fig:results-summary}(b), we randomly selected 320 testing images and computed the average SSIM scores. The proposed system outperforms the 256-QAM system in the low SNR regime (e.g., 0 to 20~dB), possibly due to the system utilization of inherent redundancies to recover noisy pixels. However, as a tradeoff, the image details tend to be lost even on high SNR regions (e.g., 40~dB), where the traditional 256-QAM system has advantages over the neural network-based methods. At IEEE ICC, we plan to take a photo of visitors and transmit the image through the real-time semantic communication system with the visual transformer.

\section{Conclusion} \label{conclusion}

In this paper, we proposed a DNN-based image semantic communication system and implemented a real-time image transmission system to demonstrate its feasibility in actual wireless environments. The proposed method exhibits promising results compared to the traditional 256-QAM based system, especially at low SNR levels. We expect these results to provide insight into the future network design for 6G.

\bibliographystyle{IEEEtran}
\bibliography{ref}

\end{document}